**A Cybersecurity MLPS Large Language Model with Multi-Path Retrieval Fusion**

*Qian Li*[1, 2], *Zhenyan Qi*[1, 2], *Liang Shen*[1, 2], *Yuan Zhang*[3], *Yifan Wan*[4], *Junyuan Ma*[5], *Yining Hu*[5*]

[1] The Third Research Institute of Ministry of Public Security, Shanghai 200031, China

[2] Shanghai Engineering Research Center of Cyber and Information Security Evaluation, Shanghai 200031, China

[3] Network and Information Center, Southeast University, Nanjing 211189, China

[4] School of Computer Science and Engineering, Southeast University, Nanjing 211189, China

[5] School of Cyber Science and Engineering, Southeast University, Nanjing 211189, China

*Corresponding author. *Email:* hyn.list@seu.edu.cn

**Abstract:**

The Multi-Level Protection Scheme (MLPS) is a foundational system in China's cybersecurity governance framework. Therefore, accurate analysis and understanding of MLPS requirements are essential. At present, MLPS analysis still relies mainly on manual interpretation of standards and rule-based tools. This makes it hard to provide stable and consistent compliance analysis in complex application scenarios. The rise of large language models has created new opportunities for making MLPS work more intelligent. However, in standards-intensive and security-sensitive scenarios, general-purpose large language models often cannot ensure controllable reasoning or complete understanding of rules.

This paper proposes a large language model framework for MLPS that integrates multiple retrieval strategies. It combines hierarchical retrieval, tree-based retrieval, and tokenization-based matching retrieval. This design helps maintain retrieval coverage while reducing the interference of irrelevant context in the reasoning process. To address the requirements of MLPS question answering for clause accuracy, conclusion traceability, and practical deployability, this paper adopts a evaluation method based on multi-dimensional weighted scoring to quantitatively assess model responses. In comparative experiments on ten typical questions, the proposed domain-specific large language model for MLPS achieved higher overall scores.

## 1. Introduction

As a foundational system in China's cybersecurity field, the Multi-Level Protection Scheme (MLPS) includes basic requirements, classification guidelines, and evaluation requirements. At present, MLPS work relies heavily on manual interpretation of standards and rule-based checking tools. In practice, this often leads to high costs, low efficiency, insufficient consistency, and limited support for continuous compliance.

In recent years, large language models (LLMs) have rapidly improved in language understanding, knowledge reasoning, and text generation. They have shown strong performance, especially in knowledge-intensive tasks. However, directly applying general-purpose LLMs to MLPS scenarios still faces several challenges. The MLPS knowledge system contains dense provisions, strong structural hierarchy, and clear context dependence. It also requires a very high level of standardization, traceability, and completeness in outputs. Without reliable external knowledge constraints, general-purpose models are prone to hallucinations and incomplete responses.

To address these issues, this paper designs and implements an intelligent question-answering and assisted analysis framework for the MLPS domain, with Retrieval-Augmented Generation (Lewis et al., 2020) as its core approach. The framework introduces customized improvements in knowledge organization, retrieval strategies, and generation constraints. As a result, it can provide highly accurate, consistent, and interpretable answers to MLPS-related questions.

### 1.1 Literature Review

Large language models have been widely applied in the cybersecurity field, including threat intelligence analysis and vulnerability detection (Ali and Ghanem, 2025; Zhang et al., 2025). They show clear advantages in improving the efficiency of security analysis, reducing manual workload, and integrating multi-source security knowledge (Divakaran and Peddinti, 2025; Xu et al., 2024). In threat intelligence analysis, current research mainly focuses on two areas: the construction of CTI datasets and evaluation benchmarks, and methods for automated analysis and detection. Della

Penna et al. (2025) built the manually annotated CTI-HAL dataset. Huang et al. (2025) developed a dataset for fake threat intelligence detection. These studies laid the foundation for cyber threat intelligence analysis. Hu et al. (2024) leveraged the capabilities of large language models in natural language understanding and knowledge reasoning, and proposed a method called LLM-TIKG. This method constructs knowledge graphs from unstructured open-source threat intelligence and effectively improves the precision of two tasks: named entity recognition and TTP classification. Bokkena (2024) introduced large language models into threat intelligence prediction and identification, and proposed an LLM-based threat prediction framework with an accuracy of 95%.

MLPS is an important foundation of China's cybersecurity assurance system. At present, research in the MLPS field mainly focuses on two areas: (1) interpretation of MLPS standards and compliance analysis (Jiang et al., 2021; Ma et al., 2019); and (2) the application of MLPS in different industry scenarios and the design of cybersecurity protection solutions based on it (Li and Qian, 2021; Zhu et al., 2020). Jiang et al. (2021) explained the background, development, and necessity of MLPS, and gave a detailed introduction to the key points and core standards of MLPS 2.0. Under the guidance of the MLPS 2.0 standards, Li and Qian (2021) proposed a general cybersecurity protection solution for the gateway station of satellite communication systems. Zhang (2025) discussed the application of MLPS in the power industry and proposed a comprehensive solution to improve cybersecurity in that sector. Zhu et al. (2020) extended MLPS to the blockchain field and evaluated and analyzed blockchain system operational data. Research on large language models in the MLPS field is still at an early stage. Existing studies mainly focus on MLPS knowledge enhancement and evaluation report inspection (Wu et al., 2025). However, current research is still exploring retrieval methods, and mature research is still lacking.

The MLPS domain is characterized by dense standards, complex hierarchies, and extremely high requirements for result reliability. To address these structural features and practical needs, this paper proposes a large language model approach that integrates workflow control, hierarchical knowledge organization, customized retrieval strategies, and generation constraint mechanisms, so as to address the limitations of existing research in this area.

### 1.2 Objectives and Contributions

This paper proposes a domain-specific large language model framework for the MLPS field based on RAG. Based on relevant regulatory documents such as *GB/T 22239-2019* (National Information Security Standardization Technical Committee (SAC/TC 260), 2019a) and *GB/T 28448-2019* (National Information Security Standardization Technical Committee (SAC/TC 260), 2019b), the framework can accurately understand user questions and produce standardized outputs, including standard references, clause identification, explanations, and difference analysis. It can support MLPS-related knowledge retrieval, consulting, and evaluation.

(1) Construction of a domain knowledge base for MLPS scenarios

A knowledge base was built around laws, regulations, and cybersecurity knowledge related to MLPS. It provides support for semantic understanding and compliance analysis by large language models in this field.

(2) Design of a multi-path hybrid retrieval strategy integrating hierarchical retrieval, tree-based retrieval, and tokenization-based matching retrieval

Given the clear hierarchical structure of MLPS standards and the strong correlation between technical terms, this paper designs a multi-path hybrid retrieval strategy. It combines hierarchical retrieval by security level, tree-based retrieval based on document title structure, and tokenization-based matching retrieval based on the BM25 algorithm. This strategy can effectively improve the accuracy of retrieval results and the coverage of retrieval for complex standards.

## 2. Problem Statements

This paper mainly addresses the application of large language models in MLPS scenarios, including intelligent standards retrieval and compliance analysis. Based on the security issue description, scenario information, and target object provided by the user, the model retrieves relevant standard content from the constructed MLPS knowledge base. On this basis, it generates analysis results that are standards-based, semantically consistent, and explainable.

## 3. Methodology

This paper proposes a domain-specific large language model framework for MLPS scenarios. The framework is constructed on a general-purpose large language model. It integrates multi-path retrieval fusion, including hierarchical retrieval, tree-based retrieval, and tokenization-based matching retrieval. The framework also incorporates MLPS domain knowledge. This design improves the reliability of the generated outputs.

### 3.1 Design of a Large Language Model Framework for the Multi-Level Protection Scheme Integrating Retrieval-Augmented Generation

In the MLPS domain, knowledge texts exhibit clear hierarchical relationships. The structures are well-defined. The terminology is highly standardized. User queries show significant variation in expression and semantic complexity. They include factual questions that correspond to specific clauses. They also include scenario-based questions that depend on business contexts.

To address these challenges, this paper proposes a large language model framework based on Retrieval-Augmented Generation. In this framework, the openPangu 7B Model (Chen et al., 2025) is used for query classification. The Qwen is used for query rewriting and answer generation. The framework improves performance in query understanding, knowledge retrieval, and answer generation.

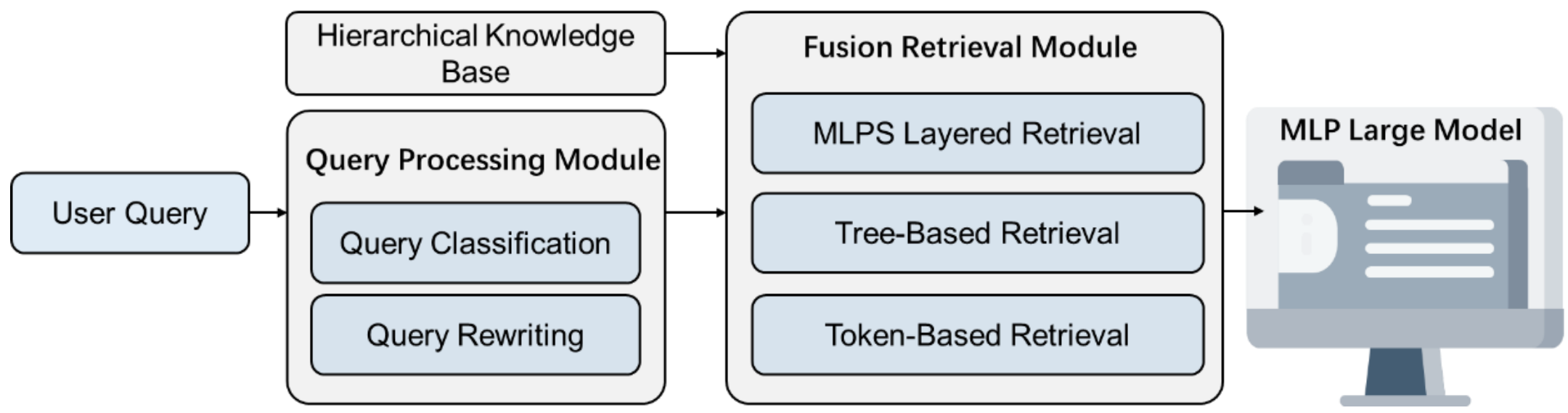

**Figure 1. MLPS-Oriented Large Language Model Framework**

(1) Query Type Identification and Abstraction Mechanism

Before retrieval and generation, the system first identifies the type of the user question and classifies it as either a factual question or a scenario-based question. This classification is based on the semantic understanding capability of the large language model. It determines whether the question depends on a specific context and whether reasoning and summarization are required.

Factual questions usually correspond to specific standard clauses, laws and regulations, or technical definitions. Their answers are deterministic and verifiable. For such questions, the main goal of the system is to improve the matching accuracy between the question and the standard texts. Scenario-based questions, however, often contain business context or hypothetical conditions, so their answers cannot be obtained directly through retrieval. The system performs semantic abstraction on such questions, reducing the influence of specific scenario descriptions and rewriting them as theoretical questions that can be aligned with normative texts. This provides a unified entry point for subsequent retrieval.

(2) Query Normalization and Semantic Alignment

To further improve retrieval performance, the system applies unified normalization to all input questions. The purpose of this process is to convert colloquial and unstructured natural language questions into retrieval queries that better match the expression style of standard documents.

The normalization process mainly includes removing colloquial expressions, eliminating semantic ambiguity, adding necessary constraints, and adopting written expressions that are more consistent with normative texts. Through this mechanism, the system can better align the structure and wording of user questions with knowledge base content such as national standards, industry specifications, and laws and regulations, without changing the original meaning of the questions.

(3) Hybrid Retrieval Module

The framework in this paper does not rely on a single retrieval method. Instead, it introduces multiple complementary retrieval strategies to improve the stability and coverage of recall. During retrieval, it combines hierarchical retrieval based on security levels, tree-based retrieval based on document title structure, and tokenization-based matching retrieval based on the BM25 algorithm to generate candidate text segments from different perspectives. These candidate results are then included in the subsequent processing pipeline and integrated through semantic reranking and authority-based constraint mechanisms, providing reliable knowledge support for the generation stage. By uniformly coordinating the hybrid retrieval strategy at the framework level, the system can improve the accuracy and stability of retrieval results while maintaining retrieval coverage.

## 3.2 Hybrid Retrieval Mechanism Integrating Multiple Strategies

In MLPS knowledge base scenarios, user queries often show strong structural features, complex semantic hierarchies, and dense technical terminology. A single retrieval strategy often has limitations in recall coverage and semantic generalization. To address this issue, this paper proposes a multi-strategy retrieval framework that integrates hierarchical retrieval, tree-based semantic retrieval, and tokenization-based matching retrieval. By combining multi-source candidate generation with unified reranking, the framework improves the stability and accuracy of retrieval in multi-document and multi-standard environments.

### 3.2.1 BM25-Based Token Matching Retrieval Strategy

BM25 is a classical sparse retrieval method that models the matching degree between query terms and document content based on term frequency (TF) and inverse document frequency (IDF). This method is highly sensitive to terminology, clause numbers, standard names, and fixed expressions, making it particularly suitable for highly standardized corpora such as security standards and regulatory texts. As a low-level exact matching retriever, BM25 builds independent indexes for each candidate document and returns relevance-ranked results based on the query terms. To avoid introducing excessive noise in multi-document scenarios, the proposed framework dynamically adjusts the number of candidates selected from each document according to document-level relevance rankings, thereby controlling candidate scale while maintaining recall. BM25 demonstrates stable performance in exact matching and effectively prevents the omission of critical clauses or key terms. However, its generalization capability for semantic paraphrasing and implicit intent is limited, making it difficult to cover natural language queries with highly diverse expressions.

### 3.2.2 Tree-Based Retrieval Strategy Based on Document Title Structure

Raptor (Sarthi et al., 2024) constructs multi-level semantic trees by leveraging the structured characteristics of documents. Leaf nodes correspond to the finest-grained original text segments, while higher-level nodes perform semantic aggregation over their child nodes through a summarization model, forming a hierarchical structure that combines both summaries and evidential content. During the retrieval stage, RAPTOR can return both leaf nodes, which provide directly citable original evidence, and routing nodes, which represent section-level or topic-level semantic summaries. Tree-based retrieval exploits document structural information, making retrieval results more consistent with human understanding of normative texts. However, its performance depends on the quality of document structures and the stability of the summarization model, and summary nodes may introduce additional noise in multi-document mixed scenarios.

### 3.2.3 Hierarchical Retrieval Strategy for Security Levels and Regulatory Clauses

The hierarchical retrieval strategy is designed to address the characteristics of security assessment and MLPS-related texts, which exhibit strong hierarchical structures and a high degree of itemization. This approach performs layered modeling of documents, such as by security level, chapter, or control point. During retrieval, it first identifies relevant documents and then conducts retrieval within each document according to its hierarchical structure. First, the large language model is used to assess document-level relevance for the query, thereby narrowing the retrieval scope. Layered retrieval is then performed within the selected documents, with the number of candidates dynamically allocated based on document relevance. Finally, a unified re-ranking model selects the most representative clause-level segments from the cross-document candidate set. Texts returned by hierarchical retrieval are typically closer to actionable items or inspection points, making them suitable for supplementing principle-level conclusions. However, this method is sensitive to the accuracy of the document selection stage. Once an incorrect document set is chosen, subsequent retrieval steps have limited ability to correct the deviation.

### 3.2.4 Fusion Mechanism of Three Retrieval Strategies

The framework proposed in this paper first merges the candidates generated by BM25, RAPTOR, and hierarchical retrieval, and then performs global ranking through deduplication and a semantic reranking model. This process ensures that the results from different retrieval strategies are compared under the same relevance standard, thereby preventing any single retriever from dominating the result distribution.

The framework also considers file priority constraints in the knowledge base. For example, national standards are assigned the highest priority. This mechanism effectively reduces the risk of conflicts caused by mixing information from multiple sources.

The hybrid retrieval strategy proposed in this paper combines tokenization-based matching, tree-based semantic modeling, and hierarchical structured retrieval. In this way, it achieves high coverage, high precision, and high reliability in complex MLPS scenarios.

## 4. Experiments and Results

To evaluate the effectiveness of the proposed MLPS-oriented large language model for cybersecurity, we constructed a question–answer dataset in collaboration with relevant user organizations and conducted experimental evaluations based on this dataset. Representative test questions are presented in the following table, covering topics such as personal information protection, identity authentication, assessment methodologies, database evaluation, system construction management, tool-based testing, security management centers, and malware protection.

**Table 1. Representative MLPS Questions**

| Representative MLPS Questions | |
|---|---|
| 1 | Analyze whether the following behaviors comply with the requirements of the MLPS standards: A financial product application requires users to actively provide information such as mobile phone numbers, login passwords, account numbers, and transaction passwords during use. In addition, the application automatically collects device information, real-time location data, contact lists, and SMS messages through permission requests, and exhibits behaviors such as frequent collection of user location information and direct uploading of users' entire contact lists. |
| 2 | For a Level-3 system of an organization, it was found during assessment that no identity authentication policy was configured on the database server. Interviews revealed that in daily operations, administrators manage the server through a bastion host that uses passwords and dynamic tokens. With respect to the assessment requirement stating that "two or more authentication technologies, such as passwords, cryptographic techniques, or biometric technologies, shall be combined for user identity authentication, and at least one of them shall be implemented using password-based technology," determine whether this requirement is satisfied and explain the reasons. |
| 3 | Briefly describe the differences and relationships between security management assessment and security technical assessment under the MLPS framework, and provide examples. |
| 4 | According to the Basic Requirements (GB/T 22239-2019), for a Level-3 protected object, which security subclasses in the secure computing environment apply to server devices? What are the contents of security auditing in the secure computing environment? Compared with Level-2 protected objects, which additional security auditing requirement is introduced for Level-3 protected objects? |
| 5 | What are common threats to databases? In MLPS 2.0 assessments, which security layer does database security assessment belong to? During the assessment process, what control points |
|  | are mainly involved in database evaluation? |
| 6 | Our company plans to build a business system with certain security requirements, and the proposed protection level is Level 2 or above, requiring MLPS assessment and protection. Please answer the following questions:<br>Question 1: From the perspective of security construction management, what aspects of work need to be completed?<br>Question 2: In security construction management, what tasks are involved in system classification and filing? |
| 7 | Please provide examples to illustrate the role of tool-based testing in the MLPS assessment process, describing its application across relevant security layers, with at least five examples. |
| 8 | Briefly describe the assessment items related to centralized management and control within the security management center. |
| 9 | What are the key inspection contents in security testing for malware protection? |
| 10 | What are the main assessment objects in security management evaluation? |

### 4.1 Data Description

The retrieval-augmented generation (RAG) framework constructed in this study takes normative documents in the fields of the Multi-Level Protection Scheme as its core knowledge sources. The knowledge base consists of 39 documents, which mainly fall into the following categories:

National standards, such as *GB/T 22239-2019* (National Information Security Standardization Technical Committee (SAC/TC 260), 2019a), which possess the highest legal and technical authority and serve as the core basis of the MLPS framework.

Industry regulations and technical guidelines, including MLPS assessment implementation guides, sector-specific technical specifications, and related supporting documents, which provide technical refinement and operational supplements to national standards.

Local standards and operational guidelines, such as implementation details, working specifications, and supplementary documents issued by local cybersecurity authorities, which reflect regional regulatory requirements and practical enforcement.

Relevant laws and regulations, including the Cybersecurity Law and the Data Security Law, which provide higher-level legal support for standards and assessment requirements.

The knowledge base documents as a whole exhibit clear hierarchical structures, dense clause numbering, and highly standardized terminology. Most documents are organized in the form of chapters, clauses, and subsections, while differences exist across documents in terms of granularity of expression, scope of applicability, and level of authority.

### 4.2 Evaluation standard

This study focuses on evaluating question answering performance in the domain of the Cybersecurity Multi-Level Protection Scheme (MLPS) by analyzing the consistency between model-generated responses and standard reference answers. Since MLPS-oriented question answering tasks are highly dependent on regulatory standards and technical specifications, the evaluation framework must simultaneously reflect compliance with cybersecurity regulations, semantic consistency with standard conclusions, and practical applicability in real-world assessment scenarios.

Considering the characteristics of MLPS assessment tasks, this paper adopts an LLM-as-a-Judge evaluation framework to conduct automatic scoring of model responses. The evaluation process compares the generated answer against the corresponding reference answer and assesses response quality from three core dimensions: completeness, accuracy, and traceability of standards. Completeness evaluates whether the generated response sufficiently covers the major control requirements, implementation measures, and assessment points described in the reference answer. Accuracy measures the semantic consistency and correctness of the generated content with respect to the standard answer. Traceability focuses on whether the response correctly references relevant MLPS standards and supporting clauses, such as GB/T 22239-2019 and GB/T 25070-2019, and whether the cited evidence properly supports the final conclusions.

To ensure quantitative and interpretable evaluation results, this study designs a weighted scoring mechanism with a total score of 10 points for each response. Specifically, completeness is assigned a weight of 5 points, accuracy is assigned 3 points, and traceability of standards is assigned 2 points. The scoring design intentionally emphasizes completeness and regulatory compliance, reflecting the strong requirement in MLPS scenarios for comprehensive security control coverage and rigorous standard alignment. During evaluation, the LLM judge outputs both a numerical score and a brief justification, enabling both quantitative comparison and qualitative analysis of model performance across different MLPS question answering tasks.

**Table 2. Evaluation Criteria**

| Dimension | Description | Weight |
|---|---|---|
| Completeness | Whether the generated response sufficiently covers the major control requirements, implementation measures, and assessment points described in the reference answer | 50% |
| Accuracy | Whether the semantic consistency and correctness of the generated content with respect to the standard answer | 30% |
| Traceability | Whether the response correctly references relevant MLPS standards and supporting clauses | 20% |

## 4.3 Results

To verify the effectiveness of the proposed framework in classified protection evaluation question-answering tasks, this study conducts comparative experiments between the proposed method and a general Retrieval-Augmented Generation (RAG) approach. During the experiments, both methods were evaluated using the same test questions and reference answers, and the generated responses were automatically assessed using an LLM-as-Judge strategy. The evaluation criteria consisted of three dimensions: completeness, accuracy, and compliance basis. Specifically, completeness accounted for 5 points and was used to measure the coverage of key points in the reference answers; accuracy accounted for 3 points and evaluated the consistency between the generated responses and the reference answers; compliance basis accounted for 2 points and assessed whether the responses correctly referenced relevant classified protection standards, such as GB/T 22239-2019 and GB/T 25070-2019. The final score ranged from 0 to 10, obtained by aggregating the three evaluation dimensions, thereby enabling a quantitative comparison of answer quality across different methods. Experimental results demonstrate that the proposed method outperforms the general RAG approach in overall evaluation scores and achieves higher scores on most test questions, indicating that the proposed framework provides better completeness, professionalism, and standard compliance in classified protection question-answering tasks.

### 4.3.1 Analysis of Experimental Results

Before evaluating the final question-answering results, the proposed framework uses openPangu 7B Model to perform hierarchical classification of user queries. In the first stage, each query is classified as either related to cybersecurity and Multi-Level Protection Scheme (MLPS) or as a general query. General queries are directed to the general-purpose response branch, whereas domain-related queries undergo a second-stage classification. The second stage further distinguishes MLPS-related queries from broader cybersecurity queries, thereby determining whether the subsequent process should invoke the domain-specific RAG pipeline.

This hierarchical classification mechanism enables the system to identify the intent and domain relevance of user queries before knowledge retrieval. During the experiment, openPangu 7B Model was able to generate classification results in accordance with the predefined categories and route the queries to the corresponding processing branches. Consequently, MLPS-related queries entered the domain-specific knowledge retrieval and answer-generation process, while unrelated or general queries were handled separately. This demonstrates the applicability of openPangu 7B Model to user-query classification and retrieval-route selection within the proposed framework.

Representative classification results further illustrate this process. For example, the query “What security requirements should a Level 3 information system satisfy?” was first identified as cybersecurity- and MLPS-related and was subsequently classified as an MLPS-specific query, thereby invoking the domain-specific RAG pipeline. The query “What are the common encryption algorithms?” was identified as cybersecurity-related but not MLPS-specific and was therefore routed to the broader cybersecurity processing branch. By contrast, a general query such as “What are the main applications of artificial intelligence?” was classified as unrelated to cybersecurity or MLPS and was directed to the general-purpose response branch. These examples show that the openPangu 7B model can distinguish among MLPS-specific, broader cybersecurity, and general queries and direct them to the corresponding processing paths.

Based on the dataset and scoring rules described above, we further evaluated the end-to-end question-answering performance by comparing the proposed framework with a general Retrieval-Augmented Generation (RAG) approach. The experimental results are presented in Table 3.

**Table 3. Experimental Results**

| ID | Proposed Framework | general Retrieval-Augmented Generation (RAG) approach |
|---|---|---|
| 1 | 7 | 7 |
| 2 | 8 | 7 |

| 3 | 9 | 7 |
|---|---|---|
| 4 | 5 | 4 |
| 5 | 6 | 7 |
| 6 | 7 | 7 |
| 7 | 5 | 7 |
| 8 | 8 | 5 |
| 9 | 7 | 5 |
| 10 | 5 | 4 |
| Average | 6.7 | 6 |

Based on the evaluation results of the 10 classified protection assessment question-answering samples shown in the table, it can be observed that the proposed framework achieves overall better performance than the general RAG approach. In terms of overall evaluation scores, the proposed framework achieved an average score of 6.7, whereas the general RAG approach obtained an average score of 6.0, representing an overall improvement of approximately 11.7%. From the perspective of individual sample performance, the proposed framework achieved higher scores than the general RAG approach on 6 out of 10 test questions, obtained equal scores on 2 samples, and scored lower on only 2 samples. These results indicate that the proposed framework is capable of generating more comprehensive and classified protection-compliant responses in most scenarios.

The experimental results indicate that the advantages of the proposed framework are mainly reflected in the following aspects.

First, in terms of referencing classified protection standards and expressing professional terminology, the proposed framework is able to conduct analyses more accurately in accordance with standards such as *GB/T 22239-2019*. For example, in questions related to identity authentication policies, secure computing environments, and personal information protection, the proposed framework can explicitly identify the corresponding control requirements and their applicable scope, thereby achieving higher scores in the dimensions of “accuracy” and “professionalism.” In contrast, although the general RAG approach can generate a certain level of summarized content, it still exhibits limitations in standard clause mapping and standardized expression.

Second, the response structure of the proposed framework is more consistent with the writing conventions commonly adopted in classified protection assessment scenarios. Most responses follow a “conclusion–analysis–basis” organizational pattern, with clear logical structure and better coverage of key points in the questions. For instance, in Question 2 regarding the “differences and relationships between security management assessment and security technical assessment,” the proposed framework not only explained the differences in assessment objects and assessment methods, but also supplemented the answer with practical assessment examples, thereby obtaining the highest score of 9 points. In comparison, although the general RAG approach provided basic conceptual explanations, it lacked further analytical elaboration and ultimately received a score of 7 points.

In addition, the proposed framework demonstrates stronger fine-grained analytical capability in complex scenarios. For example, in questions related to database security and identity authentication strategies, the framework is able to perform targeted analysis based on actual system components rather than remaining at the level of general security risk descriptions. This indicates that the framework possesses better domain adaptation capability in knowledge retrieval and contextual organization.

However, the experimental results also reveal certain limitations of the proposed framework on some questions. For example, in Question 7, the proposed framework received a score of 5, whereas the general RAG approach achieved a score of 7. This suggests that, in certain scenarios, the proposed framework may suffer from incomplete information coverage or omission of key points. These findings further indicate that the current framework still has room for improvement in terms of stability when handling long-tail knowledge and complex semantic scenarios.

### 4.3.2 Analysis of Representative Questions

(1) High-Scoring Case Analysis

Question 3, concerning the “differences and relationships between classified protection security management assessment and security technical assessment,” represents a typical strength case of the proposed framework. The proposed framework achieved a score of 9, whereas the general RAG approach obtained a score of 7.

From the generated responses, the proposed framework first provided a clear conclusion, indicating that both are essential components of classified protection assessment. It then conducted a comparative analysis from two perspectives, namely “assessment objects” and “assessment methods,” and further supplemented the response with practical assessment examples, resulting in strong completeness and professionalism. In addition, the response structure was clear and consistent with the writing conventions commonly adopted in classified protection assessment reports, which contributed to the higher evaluation score.

In contrast, although the general RAG approach was able to provide basic conceptual explanations, its content was relatively generalized and lacked in-depth analysis regarding the "relationship" and "practical application scenarios" of the two assessment types. Consequently, it was slightly inferior to the proposed framework in terms of completeness and professionalism.

(2) Compliance-Oriented Case Analysis

In Question 2 regarding the "identity authentication strategy of a Level-3 system," the proposed framework achieved a score of 8, while the general RAG approach received a score of 7.

The proposed framework was able to clearly distinguish between two types of objects, namely the "bastion host" and the "database server," and identified that the absence of an identity authentication strategy on the database server failed to satisfy the assessment requirements. This demonstrates stronger capability in differentiating assessment objects. Furthermore, the generated response incorporated specific control measures such as "password complexity policies" and "identity authentication mechanisms," making the analysis more consistent with the practical logic of classified protection assessment.

By comparison, although the general RAG approach also identified certain deficiencies, its analysis was more oriented toward general security recommendations and lacked sufficient differentiated analysis across different devices and assessment objects. As a result, its evaluation score was slightly lower.

(3) Limitation Case Analysis

Question 7, concerning the "role of tool testing in the classified protection assessment process," represents a relatively weak case for the proposed framework. In this question, the proposed framework received a score of 5, whereas the general RAG approach achieved a score of 7.

According to the evaluation results, the proposed framework may have suffered from insufficient coverage of key knowledge points in this question, which negatively affected the completeness of the generated response. In contrast, although the general RAG approach demonstrated limited professional depth, its generated content covered a relatively broader range of information, thereby obtaining a higher overall score.

This case indicates that, although the proposed framework exhibits advantages in professional compliance and structured expression, further improvements are still needed in terms of knowledge coverage capability and response stability in certain scenarios, so as to avoid performance degradation caused by omission of critical information.

### 4.3.3 Visualization Interface

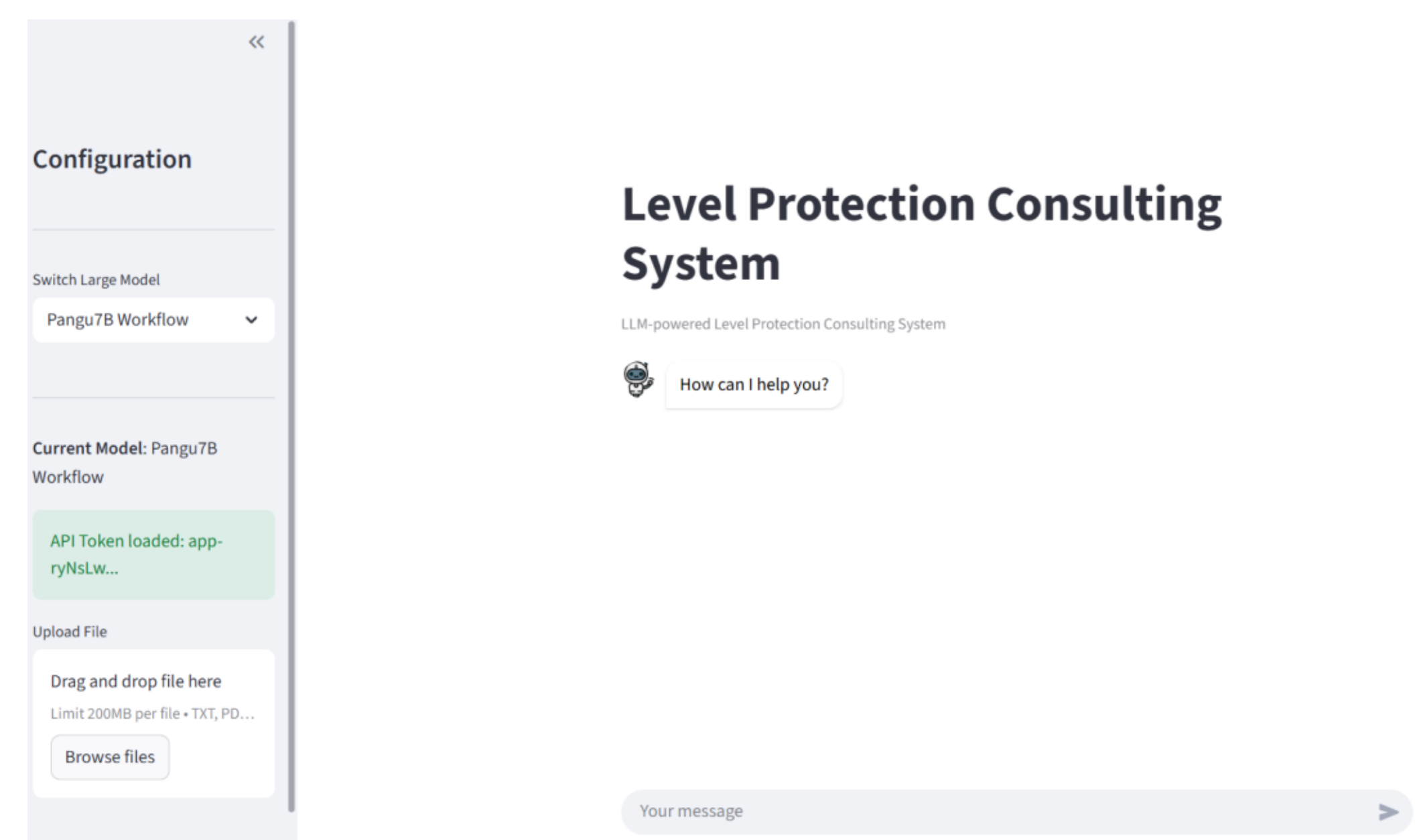

**Figure 2. Visualization Interface**

During the experimental validation process, a supporting visualized interactive interface was implemented. The front end is developed based on the React framework to support model experimentation and result presentation. The system supports multi-turn textual interactions, enabling continuous question answering with the model while maintaining contextual consistency. It also provides dynamic model switching functionality, offering a unified interaction interface for comparative experiments across different models or workflow configurations. In addition, the interface supports uploading and parsing files in formats such as PDF and TXT, allowing model evaluation to cover usage scenarios that are closer to real-world MLPS practices.

## 5. Conclusion and Discussion

This paper proposes an MLPS large language model framework integrating multiple retrieval strategies. It is designed for MLPS scenarios, where standard clauses are dense. Knowledge hierarchies are complex. Compliance analysis often depends on human experience. Through a systematic architectural design, the framework enables structured reasoning for MLPS-related question answering and analysis tasks.

In terms of knowledge support, this paper builds a domain knowledge base around MLPS-related standards and detailed specifications. Standard texts exhibit clear hierarchical relationships. Clause structures are complex. Terminology is standardized. Context dependence is strong. The knowledge content is reorganized, segmented, and restructured. This design supports effective domain knowledge enhancement.

In terms of method design, this paper adopts different models for different tasks. The framework integrates multi-path retrieval fusion, including hierarchical retrieval, tree-based retrieval, and tokenization-based matching retrieval. This design reduces the interference of irrelevant knowledge. It also maintains recall. As a result, it improves the standardization and consistency of outputs.

In terms of experimental design, this paper constructs a data-question set based on practical MLPS scenarios. The dataset covers multiple task types, including standards understanding, control point identification, clause matching, and rectification analysis. It provides a comprehensive evaluation of model performance in MLPS scenarios. The experimental results show that the proposed framework performs well in answer structure stability, completeness of control point coverage, and the ability to cite relevant clauses.

Overall, the proposed MLPS large language model framework provides a feasible and controllable technical path for intelligent MLPS applications. It also demonstrates the effectiveness of combining domain-specific retrieval strategies with large language models in MLPS scenarios.

## Acknowledgment

This study is supported by the projects TC20250926046 and KFKT2024-009.